\pdfoutput=1
\documentclass[12pt,
				a4paper,
				twoside,
				appendixprefix,
				notitlepage,
				BCOR8mm]{scrartcl}

\usepackage{color}
\usepackage{pdfpages}
\usepackage{pdflscape}
\usepackage{graphicx}
\usepackage{tabularx}
\usepackage{booktabs} 
\usepackage{authblk}

\begin{document}
\thispagestyle{empty}
\pagenumbering{gobble}

\makeatletter
\let\@fnsymbol\@roman
\makeatother


\titlehead{

\vspace{1cm}


}




\title{\Large Electron Sources for Future Lightsources,
Summary and Conclusions for the Activities during FLS 2012}


\author[1]{\normalsize C. Boulware}
\author[2]{\normalsize J. Corlett}
\author[3]{\normalsize K. Harday}
\author[4]{\normalsize F. Hannon}
\author[4]{\normalsize C. Hernandez-Garcia\thanks{chgarcia@jlab.org}}
\author[5]{\normalsize T. Kamps\thanks{kamps@helmholtz-berlin.de}}
\author[6]{\normalsize M. Krasilnikov}
\author[7]{\normalsize B. Militsyn}
\author[5]{\normalsize T. Quast}
\author[2]{\normalsize F. Sannibale}
\author[8]{\normalsize J. Teichert}

\affil[1]{\normalsize Niowave, USA} 
\affil[2]{\normalsize Lawrence Berkeley National Laboratory, USA}
\affil[3]{\normalsize Argonne National Laboratory}
\affil[4]{\normalsize Jefferson Laboratory, USA}
\affil[5]{\normalsize Helmholtz-Zentrum Berlin, Germany}
\affil[6]{\normalsize Deutsches Elektronen-Synchrotron, Germany}
\affil[7]{\normalsize Accelerator Science and Technology Center, UK}
\affil[8]{\normalsize Helmholtz-Zentrum Dresden-Rossendorf, Germany}

\date{\small 16.07.2012
}










\maketitle
\begin{abstract}
This paper summarizes the discussions, presentations, and activity of the Future Lightsources Workshop 2012 (FLS 2012) working group dedicated to Electron Sources. The focus of the working group was to discuss concepts and technologies that might enable much higher peak and average brightness from electron beam sources. Furthermore the working group was asked to consider methods to greatly improve the robustness of operation and lower the costs of providing electrons.
\end{abstract}
%
\clearpage


\clearpage
This page is intentionally left blank.



\section*{Supplementary material}

# ELECTRON SOURCES FOR FUTURE LIGHTSOURCES, SUMMARY AND CONCLUSIONS FOR THE ACTIVITIES DURING FLS 2012

C. Boulware (Niowave), J. Corlett (LBNL), K. Harkay (ANL), F. Hannon (JLAB), C. Hernandez-Garcia (JLAB), T. Kamps (HZB), M. Krasilnikov (DESY), B. Militsyn (ASTEC), T. Quast (HZB), F. Sannibale (LBNL), J. Teichert (HZDR)

*Abstract*

This paper summarizes the discussions, presentations, and activity of the FLS 2012 working group dedicated to Electron Sources. The focus of the working group was to discuss concepts and technologies that might enable much higher peak and average brightness from electron beam sources. Furthermore the working group was asked to consider methods to greatly improve the robustness of operation and lower the costs of providing electrons.

## MOTIVATION AND CHARGE TO THE WORKING GROUP

We are now living in an exciting period for future light sources, as the next generation of accelerator driven lightsources are being proposed, constructed and upgraded. The science requirements are driving the design of new lightsources, and are pushing the boundaries of our current technology. The charge given to the working group was to identify concepts and technologies that might enable much higher peak and average brightness from electron beam sources. Furthermore the working group was asked to consider methods to greatly improve the robustness of operation and lower the costs of providing electrons.

### *Injector Requirements*

For the purpose of the workshop the types of lightsources were separated in four different categories; Energy Recovery Linacs (ERL), Free Electron Lasers (FEL), Storage Rings (SR), and Compact Sources (CS). In the case of storage ring based lightsources, the injector chain typically contains a booster/accumulator ring that effectively decouples the gun performance from the main storage ring and that dramatically relaxes the requirements on the electron source. However, future lightsources might require short storage time and frequent injection for topup operation. This would mean constant uptime, high reliability and low losses for the injector. A category apart is represented by the Compact Sources, especially the plasma wakefield accelerators, where the electrons are generated in the plasma bucket itself. There will be more overlap once plasma wakefield accelerators with external electron beam injection is considered.

ERL and FEL based lightsources both require high brightness electron beams from the injector, nevertheless, lightsources at relatively long wavelengths (IR, VIS, and UV) are less demanding in terms of beam quality with respect to their shorter wavelengths (EUV to X-ray) counterparts. ERLs can potentially drive both storage ring replacement (SRR) and FEL-class lightsources. In this supergroup of lightsources, the requirements for the gun performance remarkably overlap. Indeed the necessary longitudinal and transverse phase space characteristics, expressed as 6D phase space brightness, is quite similar for both ERL-SRRs and FEL applications. The main injector requirements for ERL-SRR and FEL class lightsources are summarized in Table1.

**Table 1. Main gun requirements for EUV/X-ray lightsources based on ERL-SSRs and FELs.**

| Parameter/Feature | Requirements |
|---|---|
| Gun beam exit energy | > 0.5 MeV |
| Normalized transverse emittance | $10^{-7}$ to $10^{-6}$ mm mrad |
| Bunch length | sub-ps to tens of ps |
| Charge per bunch | Few tens of pC to 1 nC |
| Repetition rate | ~100 Hz to 1 GHz and more |
| Unwanted/wanted beam ratio | Down to $10^{-6}$ |
| Cathode launch field | > 10 MV/m |
| Compatibility | Operation of high QE photocathode inside accelerating volume |
| Reliability | Long uptime (days to months) with short service breaks |
| Availability | Quick setup of system for different modes of operation |

The bunch charge and average current can represent discriminating factors. The FEL process asks for high peak current and brightness before lasing can start. ERLs as SRR are required to produce average currents in the 100 mA range with repetition rates in the GHz range. Existing and proposed FEL facilities can be separated in low (up to 10 kHz) and high repetition rate. There is a strong desire to operate future FEL facilities in a continuous operation mode.

The electron beam quality of high brightness guns is not only dependent on the launch field and gradient inside the gun but on the properties of the photocathode itself. The contribution of the cathode intrinsic (or thermal) emittance to the normalized beam emittance can play an important role, especially when operating in the low charge (few tens of pC per bunch) regime or with the blow-out operation mode. In such a situation, the space charge effects can be effectively controlled and the final normalized emittance is dominated by the cathode

intrinsic contribution. There has been remarkably progress in photocathode R&D in recent years, mainly driven by a strong collaborative, interdisciplinary approach [M1].

*Organization of the Working Group*

We organized the working group activity in sessions dedicated to specific developments and challenges to discuss

- The state-of-the-art in NCRF and SRF guns as driver for FEL class applications to evaluate what concepts are being considered to push the envelope for FELs.
- Status and ideas for ERL class guns with SRF and DC acceleration schemes.
- Electron sources for direct beam experiments like ultra-fast electron diffraction.
- New ideas and methods for photocathodes, drive laser development and control of unwanted beam generation.

The following sections summarize the discussions and findings of the working group sessions and attempt to capture the knowledge of the community and the direction the field will be exploring in the future.

## NORMAL-CONDUCTING RF GUNS

Normal-conducting RF guns based on photoemission cathodes are one of the enabling technologies contributing to the success of short-wavelength free-electron lasers such as FLASH and LCLS. A number of ongoing projects are pushing the limits of these designs for both present and next generation free-electron laser lightsources and for other applications such as ultra-short pulse electron diffraction. Present and planned research is addressing attainable peak and average electron beam brightness and the reliability and stability of these electron sources. These developments are proceeding in parallel with photocathode research.

*PhotoInjector Test Facility, Zeuthen (DESY)*

The photo injector test facility at DESY, Zeuthen site (PITZ) is optimizing high brightness electron guns for superconducting-linac-based FELs like FLASH and the European XFEL. The main focus is on the production of short electron bunches with small transverse emittance. The group celebrated 10 years of photoelectrons at PITZ in January 2012, concluding a decade of permanent machine improvement including all the accelerator components (rf gun cavity, booster, photocathode laser system). At the same time the emittance measurement procedure is under continuous improvement in order to measure as much fraction of the electron beam transverse phase space as possible resulting in detailed experimental phase space reconstruction.

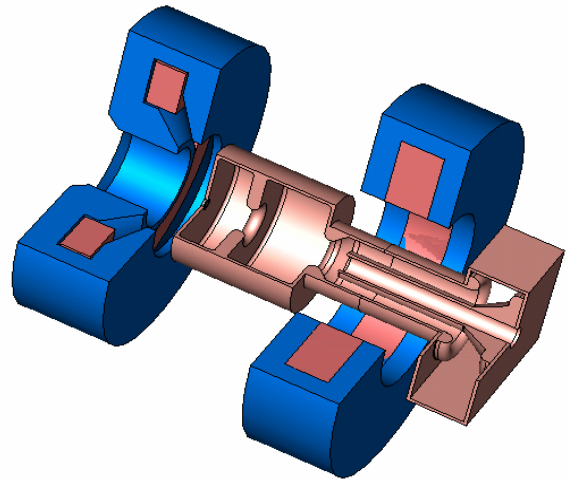

**Figure 1. The PITZ RF gun, showing the one and a half accelerating cells and the coaxial RF coupler. The main and bucking solenoids are shown with the yokes in blue.**

In 2009 specs for the European XFEL photo injector have been demonstrated for the nominal 1 nC bunch charge (the normalized rms xy-emittance of ~0.9 mm mrad has been measured) [NCRF1], the emittance has also been optimized for bunch charges of 0.1 and 0.25 nC.

**Table 2. Summary of parameters for various normal-conducting RF guns**

|  | PITZ | APEX | EBTF |
|---|---|---|---|
| operating frequency | 1.3 GHz | 186 MHz | 2.9985 GHz |
| beam energy at gun exit | 6 MeV | 0.75 MeV | 6 MeV |
| peak cathode field | 60 MV/m | 20 MV/m | 100 MV/m |
| normalized emittance at given bunch charge | 0.7 mm mrad at 1 nC (data) | 1 mm mrad at 1 nC (sims) | 0.1 to 2 mm mrad at 2 to 20 pC (sims) |
| bunch charge | up to 2 nC | up to 1 nC | up to 250 pC |
| number of cells | 1.5 | 1 | 2.5 |
| repetition rate | 10 Hz | 1 MHz (RF operates CW) | 10 Hz |

The major upgrade before the 2011 run period was dedicated to the stability of the rf gun launch phase. Installation and commissioning of an in-vacuum 10-MW directional coupler including rf controls at PITZ resulted in efficient feedback system on the rf gun phase and amplitude which improved the shot-to-shot phase stability in an order of the magnitude. This and several others machine improvements resulted in further reduction of

measured emittance (~0.7 mm mrad at 1 nC bunches) surpassing the European XFEL photo injector specs. Besides the shot-to-shot phase stability the phase homogeneity within the pulse train has been significantly improved resulting in more identical electron pulses within the long pulse trains used for emittance optimization at lower charge levels. This significantly reduced the optimum emittance number for a bunch charge of 0.1 nC (-30%, down to 0.21 mm mrad) and enabled the emittance optimization for 20 pC bunches.

Also emittance measurements for 2 nC bunch charge have been performed. Optimized measured emittance (100% rms geometric mean $\varepsilon_{xy}$): $\varepsilon$(20pC)=0.12 mm mrad; $\varepsilon$(100pC)=0.21 mm mrad; $\varepsilon$(250pC)=0.18 mm mrad; $\varepsilon$(1nC)=0.70 mm mrad; $\varepsilon$(2nC)=1.25 mm mrad. For chosen measurement conditions (fixed cathode laser pulse length) the measured emittance is close to be a linear function of the bunch charge for high bunch charges and close to the square root dependences for low bunch charges. Experimental emittance values are in a rather good agreement with expectations from beam dynamics simulations whereas the experimentally obtained optimum machine parameters deviate from the simulated ones. Further studies on improvement of the theoretical understanding of the photo-injector physics are on-going at PITZ as well as further experimental machine optimization in order to provide more flexibility for further beam manipulation in the linac and also to study the stability of the high brightness electron source performance [NCRF2].

*Advanced Photoinjector Experiment (LBNL)*

The Advanced Photo-injector Experiment (APEX) is an electron injector based on a normal-conducting (NC) continuous-wave (CW) RF photo-gun under construction at the Lawrence Berkeley National Laboratory (LBNL).

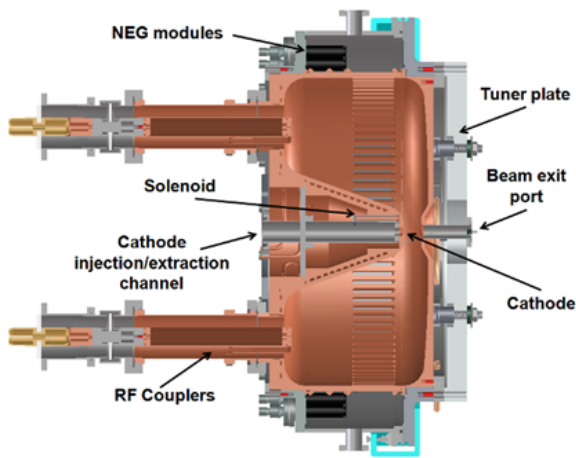

**Figure 2. The APEX VHF gun cross-section.**

The project is part of the R&D activities promoting the development of the Next Generation Light Source (NGLS), a soft x-ray light source based on an array of independently tunable FELs. The NGLS design addresses the interest of a large scientific community in the extreme ultra violet (XUV) and soft x-rays with photon energies ranging from about 10 eV to few keV requiring extremely high brightness electron sources at repetition rates as high as ~ 100 kHz per beamline.

Particularly challenging are the requirements for the electron injector to operate in the NGLS, which must deliver beams at MHz repetition rate with the required high brightness over a broad range of charge per bunch. Such an injector presently does not exist.

APEX has been designed to address such a need. APEX electron photo-gun is based on reliable and mature RF technology. The core of the gun is a NC copper RF cavity operating in CW mode in the VHF band at 186 MHz (7th sub-harmonic of 1.3 GHz). Figure 1 shows a cross section of the VHF gun with the main components, while Table 2 includes the some gun parameters.

The two major goals targeted by the gun design were the CW operation capability at high accelerating gradient at the cathode, and the high vacuum performance required to operate with sensitive high quantum efficiency (QE) semiconductor photo-cathodes. The relatively low RF frequency choice allowed addressing both of these needs. Indeed, the larger resonating structure associated with the VHF frequency decreased the heat load on the cavity wall at a level small enough to permit CW operation with conventional cooling techniques. Additionally, the long wavelength allowed opening significantly large slots on the cavity walls with negligible field distortion and creating the high conductance vacuum path required by the low pressure operation.

APEX in its final phase will include a 30 MeV linac and a suite of beam diagnostics for full 6D characterization of the electron beam phase space.

At the present time, the gun has been installed in the bunker together with a diagnostic beamline for the characterization of different photocathode performance. Figure 3 shows the APEX installed inside the bunker.

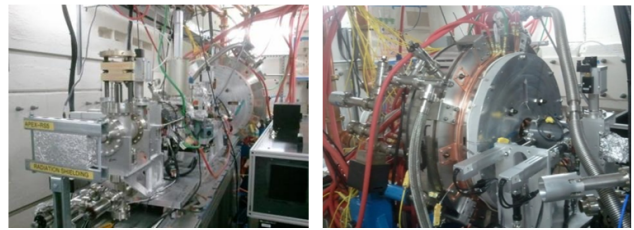

**Figure 3. The APEX in its present configuration installed inside the test area at LBNL.**

Milestones achieved so far include: full conditioning of the gun cavity at the maximum RF power; generation of first photo-emitted electron beam from a temporary metal cathode; measurement of the beam energy, confirming the design energy for the nominal power [NCRF3].

The APEX program will now continue with the test of several photocathodes, including $CsK_2Sb$, $Cs_2Te$ and diamond amplified metal cathode (in collaboration with BNL). By the end of the year, the electron beam

diagnostic suite will be installed and a full characterization of the beam performance at the target energy will follow. Finally, in late 2013, the linac will be installed and the brightness performance of APEX will be characterized.

*Electron Beam Test Facility (Daresbury)*

The Electron Beam Test Facility (EBTF) at Daresbury has the objective to provide a suite of accelerator testing facilities which can be utilized in partnership with industry, academic and scientific collaborators [NCRF4]. The high performance and flexible injector facility will comprise an RF gun, associated RF power systems, beam diagnostics and manipulators, a high power photo-injector drive laser and associated enclosures. The EBTF will serve as the front end for ongoing project CLARA, a short-wavelength seeded free-electron laser. It is also intended as a source of electrons for industrial and scientific users, to explore operation with ultra short bunches, experiments with electron diffraction, and to support a photocathode research program.

The current plan for the development of this facility involves the purchase the majority of the equipment in financial year 2011/12, with construction in 2012. First electrons from the facility are planned to be delivered in December 2012

The photocathode-driven RF gun originally designed for the ALPHA-X project [NCRF5] operates at a resonant frequency of 2998.5 MHz (S band). The design will have 4-6.5 MeV energy at the gun exit and is intended to accelerate electron bunches up to a charge of 250 pC.

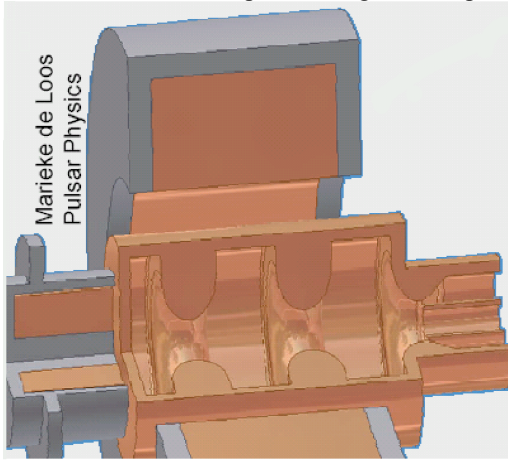

**Figure 4. The EBTF RF gun showing the two and a half accelerating cells surrounded by the main solenoid.**

The photocathodes research is underway with studies on metal photocathodes, chosen for their prompt emission in the fs-regime and their robustness under RF gun vacuum conditions. A surface-science facility for the characterization of photocathodes has been commissioned as part of the facility, to study reliable and reproducible sample preparation procedures, measure quantum efficiency of cathodes, identify effects of surface roughness and various residual gas species on the cathode performance, and investigate the impact of the back stream bombardment on the photocathode lifetime. A photocathode transport system will also be designed and its effectiveness studied.

# SUPERCONDUCTING RF GUNS

Recent years have seen significant progress with superconducting RF electron guns, which naturally scale to high duty factor because of the low losses in the cavity walls. Several programs are underway and planned to use these devices as drivers for high-repetition-rate FELs.

**Table 3. Summary of parameters for three super-conducting RF guns**

|  | ELBE | UW | HZB |
|---|---|---|---|
| operating frequency | 1.3 GHz | 200 MHz | 1.3 GHz |
| beam energy at gun exit | 9 MeV | 4 MeV | 2 MeV |
| peak cathode field | 50 MV/m | 40 MV/m | 25 MV/m |
| normalized emittance at given bunch charge | 3 mm mrad at 77 pC (data) | 0.8 mm mrad at 200 pC (sims) | 1 mm mrad at 77 pC (sims) |
| bunch charge | up to 1 nC | 200 pC | 77 pC |
| number of cells | 3.5 | 1 | 1.5 |
| repetition rate | 500 kHz, 13 MHz | MHz regime | 54 MHz 1.3 GHz |

Challenges being addressed include the integration of high quantum efficiency photo-cathodes with the cryogenic environment, and control of effects such as multipacting and field emission, which can limit the performance of such guns. Progress towards higher average currents is also being driven by developments in high-repetition-rate ultraviolet laser for driving the photocathode.

*University of Wisconsin SRF electron gun*

The University of Wisconsin, UW, is building an SRF electron gun utilizing quarter wave resonator geometry at a design frequency of 199.6 MHz [SRF1].

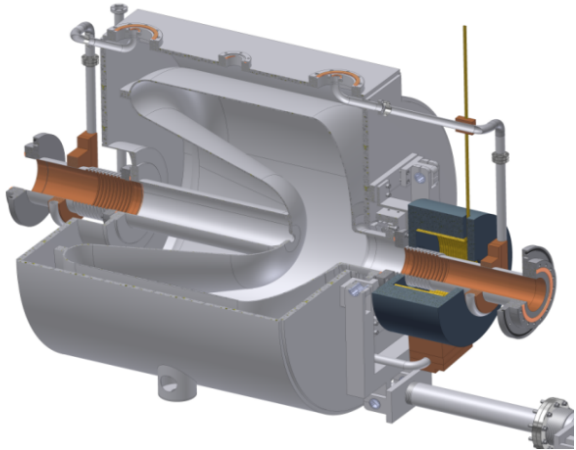

**Figure 5. The University of Wisconsin SRF quarter-wave gun inside its liquid helium vessel, with the superconducting solenoid at right.**

The gun has been modelled for a 200 pC bunch to produce an electron beam with less than 1 micron normalized transverse emittance at the design peak cathode field of 40 MV/m.

The cavity has been fabricated and tested at Niowave Inc. The measured low level Q0 is 3E9 in agreement with the Superfish runs of the cavity. The cavity was tested at Niowave up to 7 MV/m in a few hours. Additional conditioning of the cavity will be done at the University after installation into the cryostat there. Toward that goal, UW has installed a 20 kW rf transmitter utilizing a digital low level rf system procured from Jefferson Lab. A commercial cathode drive laser capable of operating at 1 kHz at 266, 532 and 800 nm has also been installed. Danfysik is fabricating a superconducting solenoid for emittance compensation, with delivery to occur in May 2012. Integration of the various parts into the cryostat is expected to occur this summer with first beam expected in Fall 2012.

## ELBE SRF Electron Gun (Helmholtz Zentrum Dresden Rossendorf)

A 3.5-cell SRF electron gun has been in operation at the Electron Linac for beams with high Brilliance and low Emittance (ELBE) machine at Helmholtz Zentrum Dresden Rossendorf (HZDR) for several years [SRF2].

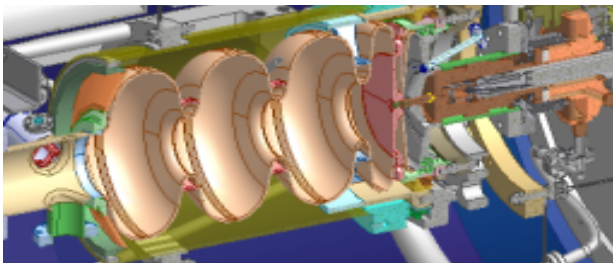

**Figure 6. The ELBE SRF gun, showing the cathode insertion region at right, the choke joint, and the three and a half accelerating cells. This gun will be used to drive the free electron laser.**

This gun is the first SRF gun to drive a free-electron laser at the ELBE-FEL facility. The gun has been designed to operate in several modes: high peak current operation for CW FELs in the IR (13 MHz, 80 pC), high bunch charge (1 nC) and low repetition rate (< 1 MHz) for pulsed neutron and positron beam production and time of flight experiments, and a low emittance, medium charge (100 pC) regime with short pulses for generation of THz radiation and x-rays by inverse Compton backscattering.

This gun has demonstrated long lifetimes with normal-conducting $Cs_2Te$ photocathodes, with operation over one year. Cathodes have emitted total charge of 35 C with quantum efficiency of 1%. No degradation in the gun cavity Q factor has been seen in 4 years, including ~2500 hours of RF operation and ~1400 hours of electron beam generation. Strong multipacting was seen at the cathode region but was defeated by a combination of curved grooves and DC biasing.

This SRF gun has been used for a Thomson backscattering experiment at ELBE, where it replaced a thermionic injector for tests. The overlap with the backscatter laser was optimized in a 10 Hz single-bunch mode. This experiment was the first demonstration of the SRF gun during a user experiment with critical needs in terms of bunch phase stability and laser-bunch synchronization.

Two new gun cavities have been fabricated in collaboration with Jefferson Lab with a slightly modified design to lower Lorentz force detuning, lower microphonics and for better cleaning and simpler cleanroom assembly. A vertical test at Jefferson Lab after Helium tank welding has demonstrated peak field of 43 MV/m in this new cavity (corresponding to 8 MeV energy gain).

Other future upgrades include a new photocathode drive laser under development at the Max Born Institute in Berlin. This laser will support the 500 kHz and 13 MHz repetition rates, allowing the gun to deliver 1 mA average current.

## BERLinPro SRF Electron Gun (Helmholtz Zentrum Berlin)

For B*ERL*inPro, an ERL test facility aiming at 100 mA average current, the electron gun needs to deliver a normalized beam emittance of better than 1mm mrad and < 20ps fwhm long electron pulses. At 1.3 GHz repetition rate, the charge per bunch is 77pC. Furthermore, the electron gun should have the flexibility to generate pulses of higher charge at lower repetition rates or shorter pulses with less charge to meet specific experimental needs.

The main challenges for such a system are achieving high average current and high beam brightness. The baseline design for BERLinPro consists of a 1.3 GHz SRF cavity equipped with a $CsK_2Sb$ cathode with a quantum efficiency of 1% at 532 nm. Given the many

challenges, the development of the photoinjector is staged. As a first step, an all superconducting 1.5 cell system (Gun0) with a superconducting Pb cathode was successfully commissioned in 2011 in the HoBiCaT facility [SRF3, SRF4]. Fig. 7 shows the cold mass of the photoinjector.

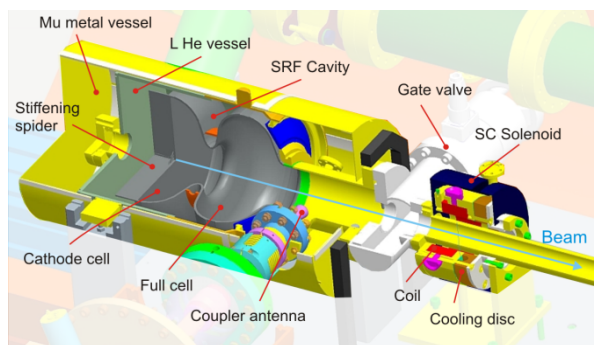

**Figure 7: Cold mass of the HZB SRF photoinjector Gun0.**

The gun demonstrated electron beam with 1.8 MeV beam energy, up 6 pC bunch charge (limited by drive laser intensity), and normalized transverse emittance of 2 mm mrad [SRF5]. Experiments have been performed to increase the quantum efficiency of the Pb cathode by laser-cleaning and to understand the process by a combined photoemission and scanning electron microscopy study.

Two additional gun prototypes will follow, the first now in the planning stage. It will include a high quantum efficiency $CsK_2Sb$ photocathode to provide the full BERLinPro bunch charge at average current of 5 mA. The second stage adds high-power RF couplers to handle the full 100 mA beam loading.

## DC PHOTOEMISSION GUNS

DC photoemission guns have recently achieved remarkable milestones despite significant technical challenges. In 2009 the Japanese Atomic Energy Agency (JAEA) DC gun demonstrated conditioning up to 550kV and stable 500kV operation [DC1]. In February 2012, the Cornell group demonstrated sustained 20 mA CW from multi-alkali photocathodes [DC2], and up to 50 mA CW from a Cs:GaAs photocathode [DC3], making the Cornell DC gun the highest CW current photoinjector in operation today.

### Cornell DC Gun

The Cornell DC photoemission gun presently operates at 350kV using a single alumina insulator [DC4]. Higher voltage is limited by field emission from the stem electrode that eventually punctures the insulator. To alleviate the problem, the Cornell team is following the JAEA segmented and shielded insulator approach. The insulator manufactured by Friatec, consisting of two assemblies each 430 mm inner diameter and 448 mm long, have been delivered to Cornell for integration to a new DC gun under development with the goal of starting high voltage conditioning in the summer of 2012. The shielding rings are made out of Copper.

### JAEA DC Gun

In 2011, the JAEA DC photoemission gun, with stem electrode coaxial to a segmented insulator with shield rings, was high voltage processing up to 510 kV, this time in its nominal configuration with the cathode electrode attached to the stem electrode, but higher voltage was limited by field emission caused by particulates falling on the electrode [DC5]. The insulator assembly consists of one single assembly 360 mm inner diameter and 750 mm long. The shielding rings are made out of titanium. After removal of particulates by wiping the cathode electrode off, and re-establishing vacuum conditions by NEG re-activation without need to re bake the gun vacuum chamber, the JAEA team was able to condition to 526kV. Albeit higher voltage is still limited by field emission, as of February 2012, stable operation at 430kV for 8 hours without any discharge has been reported. With vacuum conditions at $8 \times 10^{-10}$ Pa ($N_2$ equivalent) in the vacuum chamber with HV on, the JAEA team demonstrated 300kV electron beam generation at 15nA at the beam dump with laser power 1.4µW @ 532nm, QE=2.5% and 5.7µA at the beam dump (limited by radiation from beam dump).

The outlook for the continuing development of the JAEA DC photoemission gun includes beam generation at >400kV and approaches for high voltage conditioning in the presence of an inert gas. The DC gun is scheduled to be installed compact ERL beamline by Oct. 2012.

### KEK DC Gun

The group at KEK is also developing a DC photoemission gun based on segmented, shielded insulators with 500kV target voltage [DC6]. The titanium vacuum chamber and insulator assembly have been fabricated. The insulator assembly consists of two assemblies, each 360 mm inner diameter and 398 mm long. The shielding rings are made out of titanium. Most of the effort so far has been focused on characterizing vacuum conditions. The out gassing rate and pumping speed of the extreme high vacuum system were measured to $Q \sim 1 \times 10^{-10}$ Pa m$^3$/s. The vacuum pumping system consists of a bakeable cryopump, a series of NEG pumps (> $1 \times 10^4$ L/s, for hydrogen) placed far below the cathode electrode, and large rough pumping system (1000 Liter/s Turbo Molecular Pump).

A summary of existing and proposed DC guns can be found in Ref. [DC5].

## PHOTOCATHODES

Ultra-bright electron sources are one of the key technologies for future light source development. In particular, photocathode R&D can potentially lead to

improved electron sources to meet the stringent electron beam requirements imposed by revolutionary light source performance. In the near-term, an important activity is to optimize the synthesis and performance of long-known photocathodes such as Cs2Te, CsKSb, and GaAs. Institutions involved in this effort include: ANL, ASTeC, BNL, HZB, HZDR, INFN, JLab, KEK, LBNL, and PITZ. A longer-term approach is to explore novel crystal systems numerically and optimize ("design") their intrinsic properties for customized electron emission, or nano-engineer surfaces to prolong photocathode lifetime and to minimize unwanted beam, for example. Institutions involved in these efforts include: ANL, ASTeC, BNL, Eindhoven, HZB, JLab, LBNL, SLAC, UCLA, and Vanderbilt. Although photocathode development was a lively topic of discussion throughout the Electron Sources Working Group sessions, this summary primarily covers the three talks in the photocathode R&D session: K. Harkay (ANL) presented "Ultrabright Designer Photocathodes," W. Hess (PNNL) presented "Surface Science for Cathode Development," and B.-K. Choi (Vanderbilt U.) presented "Diamond Field-Emission Cathodes as High-Brightness Electron Sources."

ANL (Harkay, Nemeth, Terdik et al.) is exploring various novel crystal structures numerically in order to "design" their properties, such as intrinsic emittance and work function, towards the goal of developing ultra-bright photocathodes. One crystal structure is based on MgO layered with Ag, which is known to reduce the work function [PC1, PC2]. Calculations indicate that the intrinsic emittance can also be reduced. Another is a class of structures known as ternary acetylides that appear to be very promising as novel photocathodes, including an acetylated Cs2Te with a much lower work function and with a QE predicted to be similar to Cs2Te [PC3]. Synthesis and characterization of these novel systems is the next required step. PNNL (W. Hess et al.) is applying expertise in surface science techniques to characterize the properties of novel photocathodes, with the goal of better understanding the emission physics. Alkali halide or metal oxide coatings strongly modify the optical and electronic properties of hybrid structures. The surface charge (chemical potential) can thereby be tuned, dramatically enhancing the QE (CsBr on Cu) or lowering the work function (MgO on Ag). Another venue to design photoemission properties is by surface plasmon excitation. This mechanism, presently under study at LBL (H. Padmore et al.) and UCLA (P. Musumeci) allows extreme local field enhancement and can be exploited to tune the photon absorption, transmission wavelength and bandwidth on a variety of metals and alloys. Finally, Vanderbilt (B. Choi, J. Jarvis, C. Brau) is pursuing field-emission cathodes as a promising alternative to photocathodes; they are rugged and eliminate the need for a laser driver. Diamond pyramid arrays up to two inches in diameter with more than million 5-nm-tips have been fabricated, and the DC current from an individual tip was measured to be ~15 µA. The simulated ultralow emittance in a gun is very promising. Both gated and un-gated arrays are being developed, with plans to install and test one in an rf gun.

R. Legg (JLab & SRC) briefly talked of a potential low-intrinsic-emittance cathode concept that takes advantage of Dirac-like electronic band features; i.e., a small Fermi surface that is tuneable by doping or alloying. A system consisting of 20 bi-layers of Bi on a flat Si wafer is being studied at SRC. J. Teichert (HZDR) presented work to understand and improve the Cs2Te cathode lifetime T. Miyajima (KEK) presented a study of thickness-controlled GaAs cathodes and its effect on tuning the emission properties.

Discussions included other important properties to consider in studies/designs/optimization of photocathodes. Highlights include: surface roughness and contribution to the intrinsic emittance and unwanted beam (via scattered light, pointed out by P. Evtushenko (Jlab)); field emission and incompatibility of surface nano-features with high accelerating field (T. Kamps); reliability and robustness in vacuum; and emission/transmission time. It may be useful to grow novel cathodes on atomically flat substrates, such as sapphire or silicon, for the purpose of measuring their fundamental properties (B. Militsyn), and strive for more optically flat substrates in guns for reducing dark current at high fields (P. Evtushenko).

*6D phase space metric*

To better guide photocathode R&D, the complete 6-D brightness requirements for all the various future light source concepts are needed. One can then compare these requirements with gun performance and determine where development is needed. Detailed survey data have been compiled on the transverse 4-D brightness (and some 6-D data) for existing guns and those in development [PC4]. One of the outcomes of the Sources WG was a proposal to compile standardized longitudinal emittance survey data and produce a table of gun 6-D properties. K. Harkay (ANL) and B. Militsyn (ASTeC) volunteered to work on this.

## DRIVE LASER TECHNOLOGY

The desired characteristics for the drive laser of photoemission sources are sub-ps stability, micron level position stability, uniform transverse and longitudinal beam profiles.

The presently available laser power in the UV (around 250 nm) is sufficient to operate at high charge (around 1 nC per bunch) with metal cathodes at lower repetition rates (several hundreds of Hz) or with PEA cathodes like Cs2Te at 1 MHz. For the high QE photocathodes emitting in the VIS to IR (like CsK2Sb and GaAs), the generation of several tens of pC bunches at GHz repetition rate is potentially achievable with the present laser technology. Regardless of the wavelength, laser pulse shaping is an important issue for a better control of the space charge effects not completely and reliably solved yet. During the workshop three levels of laser pulse shaping have been compared with regards to performance benefits and

associated risks. The outcome of the discussion is summarized in Table 4.

*Impact of laser pulse shaping*

Beam dynamics simulations with a FEL-class gun scenario show that for 1 nC and 100 pC bunch charge a longitudinal flattop profile has 40% improvement in terms of projected emittance and 60% for 3D ellipsoidal over the emittance from a Gaussian pulse. The amount of beam halo is also reduced for the flattop and ellipsoidal pulse shapes. In case of the ellipsoidal shaping the electron beam quality is less sensitive to gun set parameter detuning.

Transverse pulse shaping has high returns in terms of beam quality for modest difficulty, longitudinal pulse shaping is more difficult and needs feedback control for reasonable returns. The most difficult is full spatio-temporal pulse shaping with yet unknown returns. Transverse pulse shaping is done most of the time by imaging an over-illuminated aperture on the cathode plane. This sounds easy in principle but requires high stability in the laserlight transport beamline as any variations in spot size and shape are transformed into intensity fluctuations resulting in bunch charge fluctuations. The principal setup of the over-illuminated aperture can improved by inclusion of an aspheric lens in front of the beam shaping aperture [L1]. The transmission is higher than for the standard setup but the sensitivity towards incoming beam fluctuations is also increased.

Longitudinal pulse shaping is done by direct space to time conversion, dazzler (acousto-optic programmable dispersive filter), spatial light modulator, and by pulse stacking. Pulse stacking systems with birefringent crystals seem to be most used solution in order to achieve longitudinal flat-top laser pulses [L2, L3, L4]. First experiments for full spatio-temporal pulse shaping have been carried out [L5] producing full 3D shaped pulses in the IR. There is still effort required to transport these pulses and to convert them into the UV. Advanced pulse shaping techniques result in advanced requirements on laser pulse diagnostics and feedback. More complicated shaping schemes utilizing non-linear effects may result in unwanted correlations between laser parameters which in turn might lead to instability. There is still R&D effort required to generate fully 3D ellipsoidal pulses and measure the benefits in terms of electron beam parameters.

Overall the impact of longitudinal pulse shaping is most on the transverse beam parameters and there mostly on the projected transverse emittance. The gain in terms of slice emittance in the central slices is small. Nonetheless a bunch with a projected emittance close to the slice emittance of the individual slices is better matched and therefore easier to transport. Un-matched tails are less populated and cause therefore less risk in terms of beam losses.

**Table 4. Comparing three levels of laser pulse shaping.**

|  | **Transverse** | **Longitudinal** | **Spatio-temporal** |
|---|---|---|---|
| Aiming at | Transverse flat-hat distribution | The beercan | 3D ellipsoidal |
| Quality gain | High | Medium | Promising |
| Biggest impact | Transverse slice emittance | Transverse and longitudinal projected emittance / reduced halo | Linear phase space / even less halo |
| Verified | Yes | Yes | In progress |
| Difficulty | Moderate | High | Very high |
| Stability | Good | Good | Problematic |
| Risk | Low | Medium | High |
| Example method | Over-illuminated aperture | Linear array of birefringent crystals | Complex array of acousto-optic modulators |
| Pro | Simple setup | Lots of shapes possible | Full control |
| Con | Reduces available laser power | Need complex diagnostics | Even more complex |

## CONCLUSION

In terms of maturity and capability of generating electron beams to meet the requirements of proposed fourth generation light sources, all three photoinjector technologies have demonstrated remarkable progress in the last couple of years.

In the low repetition rate and high brightness regime, normal-conducting RF guns are one of the enabling technologies contributing to the success of short-wavelength free-electron lasers such as FLASH and LCLS. Evolving this technology towards high repetition rate, the APEX quarter-wave normal conducting RF gun has generated photo-emitted electrons from a temporary metal cathode. With continuing development of multi-alkali cathodes, the LBL team expects to generate 1 mA CW beam with this NCRF gun.

Approaching the 100 mA CW requirement of some energy recovery linac based X-ray sources, the DC gun at

Cornell is now the highest CW current photoinjector in operation at 50 mA CW with GaAs photocathode, breaking the 32 mA CW held by the Boeing NCRF photoinjector since 1991. After demonstrating 500kV stable operation with the stem electrode co-axial to a shielded insulator, the JAEA DC gun has now achieved 526kV with the photocathode electrode in place, albeit ability to sustain this voltage is being hampered by field emission. Field emission from RF structures and DC electrodes was recognized as a common technical issue for all of the three gun technologies. Initial conversations about addressing field emission as a sub-topic in coming workshops took place amongst the working group participants.

SRF gun technology is quickly approaching performance levels already achieved with the NCRF and DC guns. For example, long lifetime of normal conducting photocathodes in SRF gun environment has been demonstrated at ELBE, along with bunch phase stability and laser-bunch synchronization. High repetition rate capability is being incorporated into two SRF guns with very different operating frequency. The B*ERL*inPro (HZB) SRF gun based on 1.3 GHz with a SC Pb cathode has already demonstrated electron beam, while University of Wisconsin has designed a quarter-wave, 199.6 MHz SRF gun built by Niowave, with first beam expected in Fall 2012.

Generating electron beams in some photoinjectors has evolved to the point of approaching the fundamental limits of the photocathode. For example, thermal emittance measurements at the LCLS photoinjector are a factor of two larger than predicted by theory, and studies by the Cornell team show the effect of photocathode surface roughness on emittance. However, there is still much left to understand.

In the FLS2010 Workshop, it was recognized that the electron beam quality of high brightness guns is not only dependent on the gradient achieved on the cathode but on the properties of the cathode. As of the FLS2012 workshop, it can be said that the injector community is entering a new era in developing photocathodes specifically engineered for a particular photoinjector (in terms of robustness and beam brightness), instead of traditional use of what is available from the semiconductor industry in terms of materials (like GaAs) and recipes (multi-alkalis). We are witnessing the first efforts for engaging with theorists, computational chemists, solid-state physicists and surface scientist towards the design of revolutionary photocathodes.

Laser 3D shaping was recognized since FLS2006 as necessary to produce high brightness beams. This topic intertwined with photocathode R&D is increasingly playing a critical role and quickly evolving as shown in Table 4.

True collaborations are developing. For example, multi-alkali photocathodes made at BNL have been transported, in vacuum, to Jefferson Lab and loaded into the CEBAF 225kV experimental inverted insulator gun; up to 20 mA DC beam has been demonstrated. Vanderbilt University is collaborating with Niowave and Fermilab to test, for the first time, diamond field emitter array cathodes in a normal conducting RF gun. Jefferson Lab and PNNL have been collaborating for the past year doing surface and bulk analyses of GaAs photocathodes that were used for years in the FEL and in the CEBAF accelerator.

# REFERENCES

[M1] D. Dowell, et al., NIM A 622 (2010) 685-697.
[NCRF1] F. Stephan, et al., PRST-AB 13, 020704 (2010).
[NCRF2] G. Vashenko, et al., Proc. of IPAC 2011.
[NCRF3] F. Sannibale, et al., Proc. of IPAC 2012.
[NCRF4] P.A. McIntosh, et al., Proc. of IPAC 2012.
[NCRF5] J. Rodier et al., proceedings of EPAC 2006.
[SRF1] R. Legg, et al., Proc. of IPAC 2012.
[SRF2] J. Teichert, et al., Proc. of FEL 2010.
[SRF3] T. Kamps, et al., Proc. of IPAC 2011.
[SRF4] A. Neumann, et al., Proc. of IPAC 2011.
[SRF5] J. Völker, et al., Proc. of IPAC 2012.
[DC1] R. Nagai, et al., RSI 81, 033304 (2010).
[DC2] L. Cultrera, et al., PRST-AB 14, 120101 (2011).
[DC3] B. Dunham, et al., Proc. of IPAC 2012.
[DC4] K. Smolenski, et al., 2009 AIP, p. 1077.
[DC5] B. Dunham, S. Benson, C. Hernandez-Garcia, N. Nishimori, T. Rao, and R. Suleiman, ERL2011 Summary of working group 1: progress with DC photoemission electron sources, 2011.
[DC6] M. Yamamoto, et al., Proc. IPAC 2011.
[PC1] K. Nemeth et al, PRL 104, 046801 (2010).
[PC2] K. Harkay et al., Proc. of PAC 2009.
[PC3] J.Z. Terdik, K. Nemeth, K. Harkay, et al., submitted for publication
[PC4] A. Arnold, J. Teichert, PRSTAB 14, 024801 (2011)
[L1] G. Klemz, et al., Proc. of FEL 2006.
[L2] H. Tomizawa, Rad. Phys. Chem. 80, 10 (2010).
[L3] A. K. Sharma et al, PRST-AB 12, 033501 (2009).
[L4] I. Will, G. Klemz, et al., Optics Express 16, 4922 14935 (2009).
[L5] Y. Li, et al., PRST-AB 12, 020702 (2009).



\end{document}